\documentclass[aps,pra,twocolumn]{revtex4}

\usepackage{graphicx}
\usepackage{pstricks}
\usepackage{bm}

\newcommand{\mb}{\mathbf}
\newcommand{\be}{\begin{equation}}
\newcommand{\ee}{\end{equation}}
\newcommand{\bwt}{\begin{widetext}}
\newcommand{\ewt}{\end{widetext}}

\begin{document}

\title{Theory of Coherent Raman Superradiance Imaging of Condensed Bose Gases}
\author{H.~Uys and P.~Meystre}
\affiliation{Department of Physics\\
The University of Arizona, Tucson, AZ, 85721}
\begin{abstract}
We model the off-resonant superradiant Raman scattering of
light from a cigar-shaped atomic Bose-Einstein condensate.
Absorption images of transmitted light serve as a direct probe of long range coherence in the condensate.
Our multimode theory is in good agreement with the time-dependent spatial
features observed in recent experiments, and the
inclusion of quantum fluctuations in the initial stages of the
superradiant emission accounts well for shot-to-shot fluctuations.
\end{abstract}
\maketitle

\section{Introduction}

In recent work, L.E. Sadler \textit{et al.} discuss the use of superradiant Raman scattering as a
probe of long range coherence in an ultra-cold bosonic gas \cite{Stamperkurn2006}.  
Their investigation is one of few experimental studies of extended sample superradiance that is not only temporally, but
also spatially resolved.

Superradiance is a well-known phenomenon \cite{Gross1982} that was
first discussed by Dicke \cite{Dicke1954} in 1954. In that process, atoms
scatter light collectively, producing a short burst of intense
radiation.  In a typical superradiance experiment with
Bose-Einstein condensates, an elongated sample of ultracold atoms
is subjected to a pump pulse of far off-resonant laser light. For an
appropriate polarization of that field, the cigar-shaped
condensate geometry results in the scattering of photons
predominantly into modes propagating along the long axis of the
sample, commonly referred to as the end-fire-modes (EFM). As the atomic cloud is
Bose-condensed, the scattering process simultaneously leads
 to the coherent amplification of the recoiling atomic
fields. Such coherent matter-wave amplification (CMA) has been
observed both in the case of Rayleigh scattering
\cite{Inouye1999a,Inouye1999b,Schneble2003a} and of Raman
scattering \cite{Schneble2004,Yoshikawa2004}, and several
theoretical descriptions of this effect have been published
\cite{Krutitsky1999,Moore1999b,Mustecaplioglu2000,Cola2004}.  

A necessary condition for superradiance to occur is \cite{Arecchi1970,Bonifacio1971,Bonifacio1975}
\begin{equation}
\label{condition}
\tau_c\ll T_2^*,
\end{equation}
where $T_2^*$ is the reciprocal inhomogeneous linewidth and $\tau_c=2/(c\rho\gamma\lambda^2)^{1/2}$ is the superradiant
cooperation time with $\rho$ the atomic density, $c$ the speed of light, $\lambda$ the wavelength and $\gamma$ the
linewidth of an isolated atom. 

An ultracold Bosonic gas at temperature $T$ with $0<T<T_c$, where $T_c$ is the
critical temperature, consists of both condensed and non-condensed phases.  As Doppler broadening for the two phases can
be dramatically different, it is possible for the appropriate choice of experimental parameters to have condition
Eq.~(\ref{condition}) satisfied for the condensed phase, but not the non-condensed phase \cite{Inouye1999a}.   Under
such circumstances light will scatter superradiantly only from the condensed portion of the gas, thus providing the
experimenter with a sensitive probe of that phase.  In Ref. \cite{Stamperkurn2006} this approach is followed to both
quantify the condensate number and study the spatial and temporal evolution of the superradiant process. It provides a
novel probe of long-range coherence which may lead to insight in the symmetry-breaking dynamics of normal- to
superfluid phase transitions.  To our knowledge only one previous study \cite{Inouye1999a} exists that
resolves spatial features in extended sample superradiance in atomic vapors.  In that case the angular pattern of the
emitted EFM was integrated over the course of a superradiant pulse and found to consist of several bright spots
(Fig. 3(a) in Ref. \cite{Inouye1999a}). 

Zobay and Nikolopoulos \cite{Zobay2005,Zobay2006} have given a detailed semiclassical analysis of the spatial
features of both the matter-wave and the optical fields in CMA
experiments based on Rayleigh scattering. They found that
propagation effects play a crucial role in the amplification
process and account for several characteristic features seen in
experiments \cite{Inouye1999a,Schneble2004}. These include the
characteristic ``X'' and fan shapes of the atomic recoil modes
corresponding to the strong and weak pulse limits respectively,
the asymmetry between forward and backward side-modes in the
strong pulse regime, and the depletion of the condensate center in
the weak pulse regime. Furthermore, they found that low
superradiant emission does not necessarily imply a small EFM field
inside the atomic sample, as a result of the scattering of EFM
photons back into the pump beam. They also demonstrated that
propagation effects lead to sub-exponential growth of the
scattered fields in contrast to the exponential growth seen in
fully quantized uniform field models, an effect reminiscent of
laser lethargy in short-wavelength optical amplifiers
\cite{Hopf1975,Hopf1976}.

An important aspect of superradiance experiments is the appearance
of large shot-to-shot fluctuations, a result of the quantum noise
that dominates the dynamics during the initial stages of the
experiment \cite{Gross1982,Haake1981,Antoniou1997,vanDorselaer1997}.
At later times the scattered fields become macroscopically
occupied and evolve in an essentially classical manner from
initial conditions determined by those quantum fluctuations. A
common strategy to treat the full evolution of superradiance
is therefore to break the problem into an initial quantum
stage followed by a classical stage. The initial stage is analyzed
primarily to determine the appropriate probability distribution of
initial conditions for the classical stage. Experimentally
observed fluctuations in the classical stage are simulated by
solving the dynamics a large number of times with initial
conditions chosen according to this probability distribution, each
simulation representing a single realization of the experiment.

The primary aim of this paper is to provide a detailed model to complement the experiments reported in Ref.
\cite{Stamperkurn2006}. We generalize the semiclassical model of 
Refs. \cite{Zobay2005, Zobay2006} to the Raman scattering case, and extend it to a multimode quantum
description that allows for a systematic treatment of the build-up
of the classical fields from quantum noise, thereby accounting for
shot-to-shot fluctuations. Whereas Refs. \cite{Zobay2005, Zobay2006} focussed on the post-pump expansion
patterns of the recoiling modes to check self-consistency of their
predictions, we investigate the time-dependent imaging of the
condensate \textit{while} undergoing superradiant emission.  We find good qualitative agreement with experimental
observations. 
 \begin{figure}
\includegraphics[angle=0, scale=0.5]{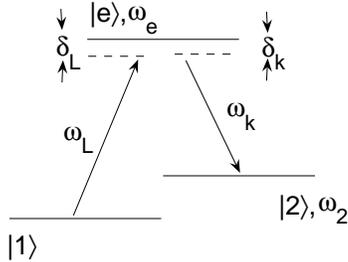}
\caption{Three-level Raman transition}\label{raman}
\end{figure}
\begin{figure}
\includegraphics[angle=0, scale=0.5]{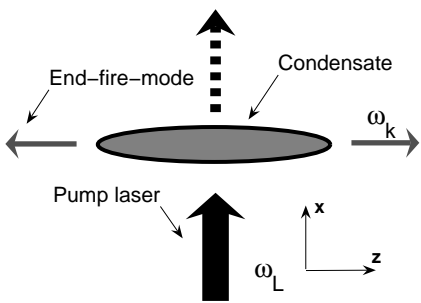}
\caption{Experimental setup.}\label{expsetup}
\end{figure}

The paper is organized as follows.  Section \ref{QF} discusses the
experiment under consideration, introduces our model, and
considers the initial stages of the evolution of the atoms and
light field, treating the scattered optical field quantum
mechanically. Section \ref{CE} turns then to the classical stage
by solving coupled Maxwell-Schr{\"o}dinger equations within the
slowly varying envelope approximation. Numerical results are
presented in Sec. \ref{NR}, and Sec. \ref{summ} is a summary and
conclusion.

\section{Initial stage and quantum fluctuations \label{QF}}

We consider a cigar-shaped Bose-Einstein condensate of width $w$
and length $L$ with a total of $N$ atoms. The ultra-cold atoms in the condensate undergo
Raman scattering between two ground states $|1\rangle$ and
$|2\rangle$ via an off-resonant excited state $|e\rangle$. The
energy of level $|1\rangle$ is taken as the zero of energy, and
the energies of the states $|2\rangle$ and $|e\rangle$ are
$\hbar\omega_2$ and $\hbar\omega_e$, respectively, see Fig. 1. We
assume that the transition $|1\rangle \rightarrow |e\rangle$ is
driven by a classical pump laser ${\bf E}_L(t)$ of frequency
$\omega_L$ and polarized along the ${\hat y}$-axis, see Fig. 2, with
    \begin{equation}
    \omega_L = \omega_e - \delta_L
    \end{equation}
while the transition $|e\rangle \rightarrow |2\rangle$ takes place via
spontaneous emission to a continuum of vacuum modes with
frequencies
    \begin{equation}
    \omega_k = \omega_e-\omega_2-\delta_k.
    \end{equation}
The total electric field is then
\begin{eqnarray}
\nonumber\mathbf{\hat E} &=& {\bf E}_L(t) +\sum_{\mb{\hat\epsilon}}\int d\mb{k}\;\mathbf{\hat E_k}\\
\nonumber &=& \mathbf{\hat{y}}\left[E_L(t) e^{i(\mb{k}_L\cdot\mb{r}-\omega_Lt)} + E_L^*(t)
e^{-i(\mb{k}_L\cdot\mb{r}-\omega_Lt)}\right]\\
\nonumber &+& \sum_{\mb{\hat \epsilon}}\int d\mb{k}\;\left[\left(\frac{\hbar\omega_\mb{k}}{2\epsilon_0
V}\right)^{\frac{1}{2}} \mb{\hat{\epsilon}_k}\hat{a}_{\mb{\hat\epsilon}\mb{k}}(t)e^{i\mb{k\cdot r}} +
h.c.\right],
\end{eqnarray}
where the creation operators
$\hat{a}^\dagger_{\mb{\hat\epsilon}\mb{k}}(t)$ obey the usual
bosonic commutation relations
\begin{equation}
\left[ \hat{a}_{\mb{\hat\epsilon}\mb{k}},
\hat{a}^\dagger_{\mb{\hat\epsilon^\prime}\mb{k^\prime}} \right] =
\delta(\mb{k}-\mb{ k^\prime})
\delta_{\mb{\hat\epsilon\hat\epsilon^\prime}}.
\end{equation}
The incident laser field envelope $E_L(t)$ is taken as
constant in amplitude during the initial stages of the
amplification process, but its full time dependence can be
accounted for in the classical stages of the evolution.

We proceed by introducing bosonic matter-field creation and
annihilation operators $\hat{\psi}^\dagger_{i}(\mathbf{r},t)$,
$(\hat{\psi}_{i}(\mathbf{r},t))$, that create (annihilate) an atom
at position $\mathbf{r}$ in electronic state $|i\rangle =
|1\rangle$, $|e\rangle$ or $|2\rangle$, with
\begin{equation}
\left[\hat{\psi}_{i}(\mathbf{r},t),\hat{\psi}^\dagger_{j}(
\mathbf{r^\prime},t)\right] =
\delta_{ij}\delta(\mathbf{r}-\mathbf{r^\prime}),
\end{equation}
in terms of which the Hamiltonian of the atom-field system is
$\hat H = \hat H_0 + \hat H_c$, with
\begin{widetext}
\begin{equation}
\label{Hfreeq} \hat H_0 = \sum_{\mb{\hat \epsilon}}\int
d\mb{k}\;\hbar
\omega(\mb{k})\hat{a}^\dagger_{\mb{\hat\epsilon}\mb{k}}(t)
\hat{a}_{\mb{\hat\epsilon}\mb{k}}(t) +\int d\mathbf{r}
\left\{\hbar
\omega_e\hat{\psi}^\dagger_{e}(\mathbf{r},t)\hat{\psi}_{e}(\mathbf{r},t)+
\hbar\omega_2\hat{\psi}^\dagger_{2}(\mathbf{r},t)\hat{\psi}_{2}(\mathbf{r},t)
\right\},
\end{equation}
\end{widetext}
and the interaction Hamiltonian
\begin{widetext}
\begin{equation}
\label{Hco}
\hat H_c = -\int d\mathbf{r} \left\{\mathbf{\hat E}\cdot \left[
\mathbf{d_1}\hat{\psi}^\dagger_{e}(\mathbf{r},t)\hat{\psi}_{1}(\mathbf{r},t)+
\mathbf{d}_2\hat{\psi}^\dagger_{e}(\mathbf{r},t)\hat{\psi}_{2}(\mathbf{r},t)
\right]
+h.c.\right\},
\end{equation}
\end{widetext}
describes the electric dipole interaction between the atoms and
the electromagnetic field, ${\bf d}_i$ being the dipole moment of
the $|i\rangle \leftrightarrow |e\rangle$ transition.

We assume that the atoms are initially in their ground state
$|1\rangle$, and that the pump laser is sufficiently far detuned
from resonance for the excited state population to remain
negligible at all times. The excited level $|e\rangle$ can then be
adiabatically eliminated in the standard fashion. Thus transforming to
slowly varying interaction picture operators
\begin{eqnarray}
\tilde{a}_{\hat{\epsilon}\mb{k}}(\mathbf{r},t) &=&\hat{a}_{\hat{\epsilon}\mb{k}}e^{i\omega_Lt}\\
\tilde{\psi}_1(\mathbf{r},t) &=& \hat{\psi}_1(\mathbf{r},t)\\
\tilde{\psi}_e(\mathbf{r},t) &=&
\hat{\psi}_e(\mathbf{r},t)e^{i\omega_et}\\
\tilde{\psi}_{2}(\mathbf{r},t) &=&
\hat{\psi}_{2}(\mathbf{r},t)e^{i\omega_2t},
\end{eqnarray}
and performing the rotating wave approximation (RWA), yields
the effective Hamiltonian
\begin{widetext}
\begin{equation}
\label{Hcq} \tilde{H}_c =-\int d\mathbf{r}\int d\mb{k}\;\left\{g(\mb{k}) \tilde{a}^\dagger_{\hat\epsilon
\mb{k}}\tilde{\psi}^\dagger_{2}(\mathbf{r},t)e^{i(\mb{k}_L-
\mb{k})\cdot\mb{r}}\tilde{\psi}_{1}(\mathbf{r},t)
e^{i(\delta_L-\delta_k)t} +h.c.\right\},
\end{equation}
\end{widetext}
where
\begin{equation} \label{gk}
g(\mb{k}) = (\hat{y}\cdot\mb{d}_1)(\hat{\mb{\epsilon}}\cdot\mb{d}_2)\sqrt{\frac{\omega_k}{2\hbar\epsilon_0V}}
\left(\frac{\delta_k+\delta_L}{\delta_k \delta_L}\right
)E_L.
\end{equation}
In Eq. (\ref{Hcq}) we have neglected higher-order scattering terms coupling modes of the scattered light
field, an approximation justified in the early stages.

The effect of the Hamiltonian $\tilde{H}$ is to transfer atoms
initially in the ground state $|1\rangle$ and with spatial wave
function $\phi_0(\mb{r})$ into the ground state $|2\rangle$ with a
momentum-conserving spatial wave function
$\phi_0(\mb{r})e^{i(\mb{k}_L-\mb{k})\cdot\mb{r}}$ via scattering
of a photon into mode $\mb{k}$. Following Ref. \cite{Moore1999b}
we therefore expand the matter-field operators into quasi-modes
according to
    \be \label{exp1} \tilde \psi_1(\mb{r}) =
    \phi_0(\mb{r})\tilde c_0
    \ee
and
    \be \label{exp2}
    \tilde \psi_2(\mb{r}) =  \int
    d\mb{q}\;\phi_0(\mb{r})e^{i\mb{q}\cdot\mb{r}}\tilde c_q,
    \ee
where $\tilde c_0$ annihilates a particle in a quasi-mode with
electronic state $|1\rangle$ and wave function
$\langle\mb{r}|0\rangle=\phi_0(\mb{r})$ and $\tilde c_q$
annihilates a particle in a quasi-mode with electronic state
$|2\rangle$ and wave function $\langle\mb{r}|\mb{q}\rangle =
\phi_0(\mb{r})e^{i\mb{q}\cdot\mb{r}}$. These quasi-modes are
nearly orthogonal provided that they are separated by an angle
    $$
    \theta_\bot \gtrsim 2K_w/|k| = \lambda/(\pi w),
    $$
where $K_w = 2/w$ is the momentum width of the condensate. In that
case the quasi-mode creation and annihilation operators $\tilde
c_q$ obey to a good approximation bosonic commutation relations
\begin{equation}
\label{cqcom}
\left[ \tilde{c}_q,\tilde{c}^\dagger_{q^\prime} \right]
\approx \delta_{q q^\prime}
\end{equation}
and are statistically independent during
the early stages of the evolution.

In terms of the quasi-modes $q$, the Hamiltonian (\ref{Hcq})
becomes
    \begin{equation}
    \label{Heff}
    H_c = \sum_\mb{q}\int
    d\mb{k} \left\{\eta(\mb{k},q)\;e^{i(\delta_L-\delta_k)t}\tilde
    a^\dagger_\mb{k} \hat c^\dagger_q \hat c_0+h.c.\right\}
    \end{equation}
where
    \be
    \label{eta}
    \eta_\mb{q}(\mb{k}) = g(\mb{k})\int
    d\mb{r}|\phi_0(\mb{r})|^2e^{i(\mb{k}_L-\mb{k}-\mb{q})
    \cdot\mb{r} }.
    \ee
We have used the normalization
condition $\int d\mb{r}\;|\phi_0(\mb{r})|^2= 1$. The Hamiltonian
(\ref{Heff}) differs from the effective Hamiltonian derived by
Moore and Meystre for the Rayleigh scattering case
\cite{Moore1999b} in that the electronic state of recoiling atomic
modes is $|2\rangle$ instead of the original ground state
$|1\rangle$ and the term oscillating with frequency
$\delta_L-\delta_k=\omega_k-(\omega_L-\omega_2)$ enforces the
coupling of the scattered modes to photons with frequency
$\omega_L-\omega_2$ instead of $\omega_L$.

With these differences in mind, we may easily generalize the
results of Ref. \cite{Moore1999b}  and determine that the quasi-modes
initially grow exponentially,
    \be \label{ct}
    \tilde c_\mb{q} =
    \exp[G_\mb{q}Nt/2]\tilde c_\mb{q}(0) +
    \int_0^td\tau \exp [G_\mb{q}N t/2]\tilde
    f_\mb{q}^\dagger(t-\tau),
    \ee
where
    \be
    G_\mb{q} = 2\pi\int d\mb{k}\;|\eta_\mb{q}(\mb{k})|^2
    \delta[\omega_k-(\omega_L-\omega_2)]
    \ee
and $\tilde f_\mb{q}(t-\tau)$ is a noise operator the
second-order correlation functions of which are given in the Markov
approximation by
\begin{eqnarray}
\langle \tilde f_\mb{q}^\dagger(t)\tilde f_\mb{q}(t^\prime)\rangle &=& 0,\\
\langle \tilde f_\mb{q}(t)\tilde f_\mb{q}^\dagger(t^\prime)\rangle &=&
G_\mb{q}N\delta(t-t^\prime).
\end{eqnarray}
In this limit, the probability $P_\mb{q}(n,t)$ of having $n$
atoms in mode $\mb{q}$ at time $t$ is that of a chaotic field,
    \be \label{pdist}
    P_\mb{q}(n,t) = \frac{1}{\bar n_\mb{q}(t)}\left
    (1+\frac{1}{\bar n_\mb{q}(t)}\right )^{-(n+1)},
    \ee
where $\bar n_\mb{q}(t)= \langle \tilde c^\dagger_\mb{q}\tilde
c_\mb{q} \rangle$ is the mean number of atoms in the quasi-mode
$q$ at time $t$.

An important feature of the linear gain factor $G_\mb{q}$ is that
it remains relatively constant for quasi-modes excited via the
scattering of photons at small angles $\theta_\mb{k}$ with respect
to the long axis of the condensate. This is the case until
$\theta_\mb{k}$ reaches the geometric angle
    $$
    \theta_g\approx\frac{w}{L}
    $$
after which the linear gain $G_\mb{q}$  falls off rapidly. The
electromagnetic modes corresponding to scattering into that angle
collectively form the EFMs, and they dominate the short-time dynamics of the system.

This suggests that we may accurately simulate the linear dynamics
of the superradiant system by considering the scattering of
photons into a finite number of modes distributed within a solid
angle $2\pi\theta_g$ around the long axis of the condensate only.
We note that while the geometric angle into which significant
scattering takes place is fixed by the aspect ratio of the
condensate, the number $m$ of independent quasi-modes depends also
on the wavelength of the optical fields involved,
    \be
    \label{Mmodes}
    m \approx \left(\frac{\theta_g}{\theta_\bot}\right)^2
    =\left(\frac{\pi w^2}{2\lambda L}\right)^2 = F^2,
    \ee
where $F$ is the Fresnel number. Typical
experiments correspond to a number of quasi-modes
$m\approx1\sim10^2$.

\section{Classical Evolution\label{CE}}

We now turn to the quantum-noise initiated classical regime that
occurs once the superradiance process is fully underway and the
scattered modes are macroscopically occupied. In typical experiments of Ref. \cite{Stamperkurn2006} the velocity of
the recoiling atoms is such that they may traverse half a condensate width over the course of a single run of the
experiment. In addition, atom-atom scattering may cause significant dephasing on the same time scale.  In this section
we therefore include both kinetic energy and mean field atom-atom collisional terms.

In the classical regime the optical field can be described by a complex field amplitude
\begin{eqnarray}
\nonumber\mathbf{E}_{cl} &=& \mathbf{E}_L+\sum_\mb{
k}\mathbf{E_k}\\\nonumber &=&  \mathbf{\hat{y}}\left[
E_L(\mathbf{r},t)e^{i(\mb{k}_L\mb{\cdot r}-\omega_Lt)} +
E_L^*(\mathbf{r},t)e^{-i(\mb{k}_L\mb{\cdot r}-\omega_Lt)}\right]\\\nonumber &+& \sum_{\mathbf{\hat{\epsilon}}\mb{
k}} \mathbf{\hat{\epsilon}}\left[E_\mb{k}(\mathbf{r},t)e^{i(\mb{k\cdot r}-\omega_kt)} +
E_\mb{k}^*(\mathbf{r},t)e^{-i(\mb{k\cdot r}-\omega_kt)}\right].
\end{eqnarray}
Following the discussion of Sec. \ref{QF} we restrict the sum over
$\mb{k}$ to the discrete set of quasi-modes in the end-fire cone
and set $\omega_k = \omega_L-\omega_2$. At this point we no longer assume that the pump field remains undepleted. As
usual, the field amplitudes $E_i(\mb{r},t)$ are assumed to be
slowly varying,
    \begin{eqnarray}
    \label{slowaprx}
    \left |\nabla E_i(\mb{r},t)\right | &\ll&
    \left |k_iE_i(\mb{r},t) \right |,\\
    \left |\frac{\partial E_i(\mb{r},t)}{\partial t}\right | &\ll&
    \left |\omega_i E_i(\mb{r},t)\right |.
    \end{eqnarray}
Eliminating adiabatically the excited electronic state of the
atoms as before by introducing the slowly varying Schr{\"o}dinger
field operators
\begin{eqnarray}
\tilde{\psi}_0(\mathbf{r},t) &=& \hat{\psi}_1(\mathbf{r},t)\\
\tilde{\psi}_{\mb{k}_L}(\mathbf{r},t) &=&
\hat{\psi}_e(\mathbf{r},t)e^{i(\omega_e + \omega_{er})t}e^{-i\mb{k}_L\cdot\mb{r}}\\
\tilde{\psi}_\mb{k}(\mathbf{r},t) &=&
\hat{\psi}_\mb{k}(\mathbf{r},t)e^{i(\omega_2+\omega_{2r})t}e^{-i(\mb{k}_L\cdot\mb{r}-\mb{k}\cdot\mb{r})},
\end{eqnarray}
where $\hbar\omega_{er} = \hbar^2k_L^2/2m$ and $\hbar\omega_{2r} =
\hbar^2(k_L^2+k^2)/2m$, results in the effective Hamiltonian 
$\tilde{H} =\tilde{H}_{0}+\tilde{H}_{aa}+\tilde{H}_c$ where
\begin{equation}
\label{Hkin}
\tilde{H}_{0} = -\int
d\mathbf{r}\left\{\sum_\mb{k}\frac{\hbar^2}{2m}\tilde{\psi}_\mb{k}^\dagger\nabla^2\tilde{\psi}_\mb{k}\right\},
\end{equation}
\begin{widetext}
\begin{equation}
\label{Haa}
\tilde{H}_{aa} = -\int
d\mathbf{r}\left\{U_{11}\tilde{\psi}_0^\dagger\tilde{\psi}_0^\dagger\tilde{\psi}_0\tilde{\psi}_0+\sum_\mb{k}\left(U_{kk}
\tilde{\psi} _\mb{k}^\dagger\tilde{\psi}_\mb{k}^\dagger\tilde{\psi}_\mb{k}\tilde{\psi}_\mb{k}+
U_{1k}\tilde{\psi}_0^\dagger\tilde{\psi}_\mb{k}^\dagger\tilde{\psi}_\mb{k}\tilde{\psi}_0\right)\right\},
\end{equation}
\begin{equation}
\label{Hcl}
\tilde{H}_c = -\int d\mathbf{r}
\sum_{ij}\frac{g_{ij}}{(1-i\gamma/\delta_L)}\tilde{\psi}^\dagger_{i}(\mathbf{r},t)\tilde{\psi}_{j}(\mathbf{r},t).
\end{equation}
\end{widetext}
Here
\begin{equation}
\label{gj}
\label{gjk}
g_{ij}=\frac{d_id_j}{3\delta_L\hbar}E_i^*(\mathbf{r},t)E_j(\mathbf{r},t),
\end{equation}
in which the factor $3$ arises from an average over all possible orientations of the dipole and
\begin{eqnarray}
\label{Uaa}
U_{11(kk)}=\frac{4\pi\hbar^2a}{m},
U_{1k}=\frac{8\pi\hbar^2}{m}\left(a+ika^2\right),
\end{eqnarray}
where $a$ is the s-wave scattering length and the indices $i,j$ run over
$-k,0,k$. The last term on the right
in Eq.~(\ref{Uaa}) is a momentum-dependent loss that has been used to account for elastic scattering losses in the
Gross-Pitaevski equation \cite{Band2000}.  It arises upon keeping the second-order term when expanding the
manybody T-matrix in powers of $k$.  For the Raman system under
consideration we need not include higher-order modes
of the form $\tilde \psi_{n{\mb k}_L+q\mb{k}}$ ($n$,$q$ being integers) in the
sum over $\mb{k}$, since the second ground state can no longer absorp pump photons \cite{Schneble2004}.

Within the slowly varying envelope approximation (SVEA) the Hamiltonian
(\ref{Hcl}) yields the matter-wave Heisenberg equations of motion
\begin{widetext}
\begin{eqnarray}
\frac{d
\Psi_0(\mathbf{r},t)}{d\tau}&=&\frac{i}{(1-i\gamma/\delta_L)}\mathcal{E}_L^*\left[\frac{d_1}{d_2}\mathcal{E}_L{
\Psi}_0+\sum_\mb{k}\mathcal{E}_\mb{k}\Psi_\mb{-k}\right]+i\sum_\mb{j}u_{1j}|\Psi_j|^2\Psi_0\label{psi1}\\
\frac{d\Psi_\mb{k}(\mathbf{r},t)}{d\tau}&=&-\mb{\kappa}\cdot\nabla
\Psi _\mb { k } +\frac{i}{(1-i\gamma/\delta_L)}\mathcal{E}_\mb{-k} ^*\left
[\mathcal{E}_L{\Psi}_0+\frac{d_2}{d_1}\sum_\mb{k}\mathcal{E}_\mb{k}\Psi_\mb{-k}\right]+i\sum_\mb{j}u_{kj}
|\Psi_j|^2\Psi_k\label{psi2}.
\end{eqnarray}
\end{widetext}
We have cast these equations in dimensionless units where the dimensionless time is 
\be
\tau = \frac{\Omega_L^2}{\delta_L}t,
\ee
in which
    \be
    \Omega_L = \frac{d E_L^{{\rm in}}}{\sqrt{3}\hbar}
    \ee
is the the Rabi frequency of the incident
laser field of amplitude $E_L^{\rm in}$ and $d=\sqrt{d_1d_2}$ .  The elecric field amplitudes are rescaled as
    \be \label{Edls}
    \mathcal{E}_i =  \frac{ E_i}{E_L^{\rm in}},
    \ee
and the matter-wave fields as
    \be
    \label{Pdls}
    \Psi_i = \frac{\tilde\psi_i}{\sqrt{\rho_c}},
    \ee
where
    \be
    \rho_c = \frac{N}{Lw^2}
    \ee
is a characteristic density. 

When performing the derivatives of the kinetic energy term, dimensional analysis along with the SVEA indicate that the
second order derivative is smaller than the first by a factor $10^{-4}$.  We have therefore retained only the
advective contribution (first term on the right in Eq.~(\ref{psi2})) with a dimensionless velocity 
\be
  \mb{\kappa} = \frac{\delta_L}{L\Omega_L^2}\frac{\hbar}{2m}(\mb{k}_L+\mb{k}).
\ee
Finally, the rescaled atom-atom scattering strength is 
\be
	u_{ij} =  \frac{\delta_L}{\Omega_L^2}U_{ij}.
\ee

The factor $\gamma/\delta_L$ in Eqs.~(\ref{psi1}) and (\ref{psi2}) is due to the inclusion of a phenomenological decay
term to account for losses due to spontaneous emission from the excited state.

The evolution of the optical field is governed by the Maxwell wave equation coupled to the macroscopic polarization of
the condensate.  Treating the pump laser as a continuous wave we find

\begin{eqnarray}
\nonumber\frac{\partial \mathcal{E}_L}{\partial \xi}&=&
\frac{i\aleph}{(1-i\gamma/\delta_L)}\Psi_0^\dagger\left\{\frac{d_1}{d_2}\mathcal{E}_L\Psi_0
+\sum_k\mathcal{E}_k\Psi_{-k}\right\}\\
\label{propagate1} & &\\
\nonumber{\rm sign}(k) \frac{\partial\mathcal{E}_k}{\partial \zeta} &=&
\frac{i\aleph}{(1-i\gamma/\delta_L)}\Psi_{-k}^\dagger\left\{\mathcal{E}_L\Psi_0
+\frac{d_2}{d_1}\sum_k\mathcal{E}_k\Psi_{-k}\right\}\\
\label{propagate2} 
\end{eqnarray}
where $\xi = x/L$, $\zeta = z/L$ and
    \be
    \aleph = \frac{\pi d^2\rho_c}{3\epsilon_0\hbar \delta_L}\frac{L}{\lambda}.
    \ee
Equations (\ref{propagate1}) and (\ref{propagate2}), together with
Eqs. (\ref{psi1}) and (\ref{psi2}), fully describe the multimode
dynamics of the slowly varying envelopes of the electric and
matter-wave fields.  They may be solved analytically in the
short-time regime as shown in Appendix A.  In this regime the fields $\Psi_{\pm k}$
obey
\be
\Psi_{k}(\mathbf{r},\tau) = \psi_{k^\prime}(0)I_0\left(2\sqrt{\tau\Delta}\right),
\ee
where $\psi_{k^\prime}(0)$ is the square root of the recoiling mode density at $\tau=0$ and $\Delta$ is a function of
$\zeta$ defined in Appendix A.  A similar result was previously obtained in Ref. \cite{Zobay2006}.

\section{Numerical Results \label{NR}}

\subsection{General considerations}

\begin{figure}
\centering
\begin{center}
\includegraphics[angle=0, scale=0.5]{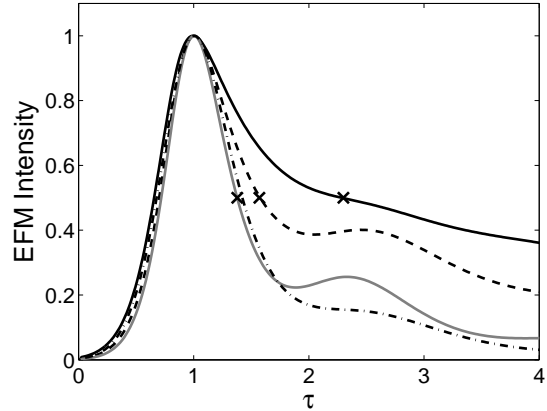}
\end{center}
\caption{Typical time evolution of the EFM intensity.  These simulations neglect dissipation, mean-field interactions
and effects due to photon recoil.  The curves correspond to: solid line $\aleph=6.0$, dashed line $\aleph=3.0$,
gray line $\aleph=1.0$ and dot-dashed line $\aleph=1.0$ with $\gamma/\delta_L=0.04$.  The x's mark points at which the
spatial field and absorption profiles are plotted in Figs. \ref{psifig}-\ref{absrp}.}\label{superfig}
\end{figure}
\begin{figure}
\centering
\begin{center}
\includegraphics[angle=0, scale=0.5]{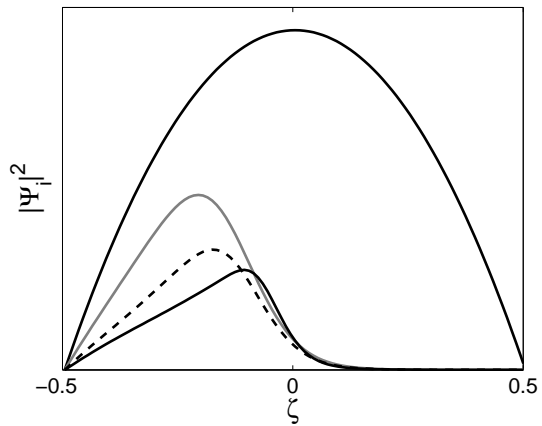}
\end{center}
\caption{Density profile of the condensate at $\tau=0$ (top solid line) and of the
atomic side mode $\Psi_{+k}$ at half the maximum EFM intensity, as indicated by x's in Fig.
\ref{superfig}: solid line $\aleph=6.0$, dashed line $\aleph=3.0$ and gray line $\aleph=1.0$. Mean-field interactions,
dissipation and effects due to photon recoil are neglected.}
\label{psifig}
\end{figure}

This section presents selected results from numerical simulations
of the onset and growth of superradiant scattering in a
cigar-shaped condensate. We assume here that all atoms in the
condensate at temperature $T=0$ are initially in the electronic
ground state $|1\rangle$, and approximate their center-of-mass
wave function by a separable Thomas-Fermi profile (in dimensionless units),
$$
\Psi_0(\mb{\mathcal{R}})=
6^{3/2}L^2/w^2\prod_i\sqrt{(\mathcal{L}_i/2L)^2-\mathcal{R}_i^2}
$$
where the product is over the rescaled spatial coordinates $\mathcal{R}_i$ and $\mathcal{L}_i$ is the
corresponding condensate width 
\footnote{
Around the center of the wavefunction this approximation has small corrections, second order
in $x_ix_j/r_ir_j$, to the true Thomas-Fermi wavefunction.}.
Close to the Thomas-Fermi radii, where the Thomas-Fermi approximation breaks down, we let the
wavefunction go to zero smoothly by matching both the function and its derivative to a Gaussian tail.  This is
necessary to prevent numerical instabilities resulting from the advective term in Eq.~(\ref{psi2}).  To remain true to
the experiments of Ref. \cite{Stamperkurn2006}, we choose the aspect ratio of the condensate to be
$w/L \approx 0.1$.

As we have seen, the amplitudes of the end-fire modes at the onset of the classical regime are stochastic
variables the values of which must be selected at random from run-to-run
consistently with the results of Sec. II. However, to set the stage for our discussion, we consider first a
simplified situation with only one left-recoiling and one right-recoiling
side-mode of equal initial amplitudes small compared to the
amplitude of $\Psi_0$, but with the same spatial structure.  We include, for the moment, only the atom-light coupling
and neglect the mean-field interaction, photon recoil and dissipation.  

\begin{figure}
\centering
\begin{center}
\includegraphics[angle=0, scale=0.5]{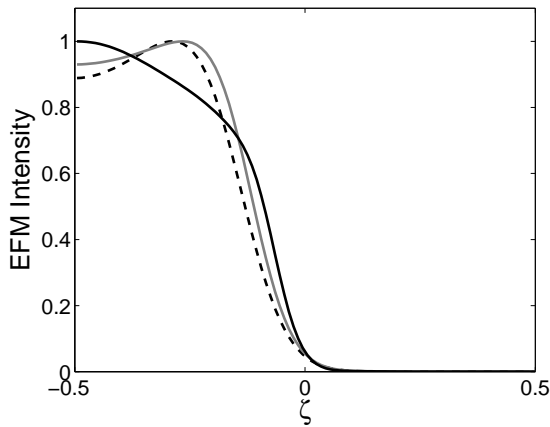}
\end{center}
\caption{Typical spatial envelopes of EFM electric fields at half maximum EFM intensity as indicated by x's in Fig.
\ref{superfig}: solid line $\aleph=6.0$, dashed line $\aleph=3.0$, gray line $\aleph=1.0$.  The peak in intensity for
$\aleph=1.0$ and $3.0$ arises due to scattering of EFM photons back into the probe beam. Mean-field interactions,
dissipation and effects due to photon recoil are neglected here.}\label{efmfig}
\end{figure}
\begin{figure}
\includegraphics[angle=0, scale=0.5]{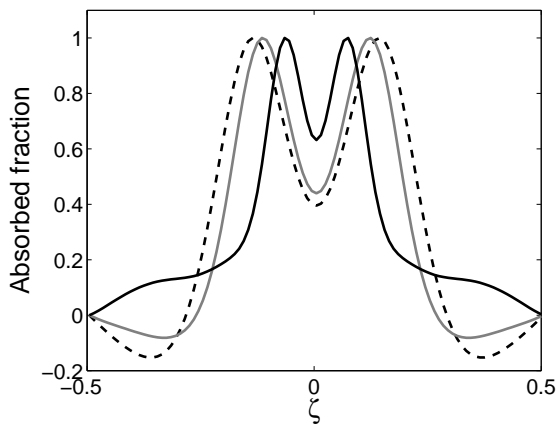}
\caption{Absorption profile of the pump pulse along the $z$ axis
at half maximum EFM intensity as indicated by x's in Fig.
\ref{superfig}: solid line $\aleph=6.0$, dashed line $\aleph=3.0$ and gray line
$\aleph=1.0$. Mean-field interactions, dissipation and effects due to photon recoil are neglected
here.}\label{absrp}
\end{figure}

Figure \ref{superfig} illustrates examples of
superradiant EFM pulse shapes, for different values of the atom-field coupling strength $\aleph$.  To facilitate
comparison we scaled the peak intensities to $1$, and the time axes so that the peaks coincide. The gray
line, corresponding to $\aleph=1.0$, demonstrates a typical main pulse followed by a secondary peak, an effect
known as ringing \cite{Gross1982}.  Increasing the coupling strength to $\aleph=3.0$ raises the secondary peak maximum
relative to the first peak, as shown by the dashed line.  The black line, for which $\aleph=6.0$, represents a
qualitatively different regime: The superradiant maximum is now followed by a slowly decaying
tail rather than ringing. Dissipation also leads to suppression of the ringing, as shown by the dot-dashed line which
is for $\gamma/\delta_L=0.04$ and $\aleph=1.0$. The appearance of ringing with increased coupling strength has
been observed
experimentally, Fig. 3(b) of Ref. \cite{Inouye1999a}.

Figures \ref{psifig} and \ref{efmfig} show spatial profiles along the long axis $z$ of the condensate,
of the atomic side mode $\Psi_{+k}$ and of the corresponding EFM field $\mathcal{E}_{-k}$. Here the same values of
$\aleph$ as in Fig.~\ref{superfig} were used: solid line $\aleph=6.0$, dashed line $\aleph=3.0$ and gray line
$\aleph=1.0$.  The EFM profiles have again been 
scaled for comparison.  All $z$-axis profiles correspond to the times indicated by the x's in Fig.~\ref{superfig}.  
In Fig. \ref{psifig} the  atomic mode $\Psi_{+k}$ grows from the left edge of
$|\Psi_0|(\zeta)^2$, a consequence of the build-up of the EFM
electric field as it moves across the
condensate, see Fig. \ref{efmfig}.  For clarity the mode $\Psi_{-k}$ is not shown. It is
simply a mirror image of $\Psi_{+k}$.  Interestingly, in the absence of
dissipation the weaker coupling strength is more efficient at scattering atoms to ground state $|2\rangle$ over a single
superradiant pulse, as evidenced by the gray line in Fig. \ref{psifig}.  Keep in mind, however, that the time
scale is slower for the weaker coupling, a feature not apparent due to the scaling of time in the figure.

For the cases $\aleph=1.0$ and $3.0$, the EFM profile peaks before reaching the edge of the condensate, see Fig.
\ref{efmfig}. During the dynamical evolution this peak
appears shortly after the first intensity maximum in Fig. \ref{superfig}.  It is a consequense of the EFM field being
scattered back into the
probe beam from atoms in ground state $|2\rangle$.  As pointed out in Ref. \cite{Zobay2006} this may cause a
strong EFM field to exist within the atomic system despite small
emission outside the sample.
\begin{figure}
\includegraphics[angle=0, scale=0.5]{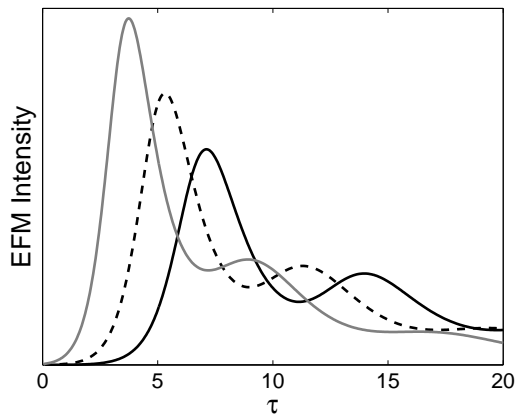}
\caption{Time evolution of EFM intensity for an initial density of the recoiling modes $10^{-6}$ (solid black line),
 $10^{-5}$ (dashed line) and  $10^{-4}$ (gray line) times smaller than the condensate density.}\label{ampfig}
\end{figure}

The fraction of absorbed laser intensity 
\begin{equation}
\nonumber
A(x,z)=\frac{|\mathcal{E}_L^{{\rm in}}|^2-|\mathcal{E}_L(x,z)|^2}{|\mathcal{E}_L^{{\rm in}}|^2}
\end{equation}
is shown in Fig. \ref{absrp} for $x\gg w$ after the laser field exits the condensate, again for the same values of
$\aleph$ as in Fig.~\ref{superfig} and at half the EFM intensity maximum. 
The two central peaks in each absorption profile appear close to the edges of the condensate early in the time
evolution and migrate towards each other.  Secondary features, such as regions of gain around the edges of the
condensate, appear shortly after the EFM pulse maximum for $\aleph=1.0$
and $3.0$.  This gain is a signature indicating that
EFM light is being scattered back into the pump field. For $\aleph=6.0$ the growth of secondary absorption peaks
at the edges of the condensate, as shown by the black line, is the dominant secondary feature rather than gain.
\begin{figure}[t]
\centering
\begin{center}
\includegraphics[angle=0, scale=0.5]{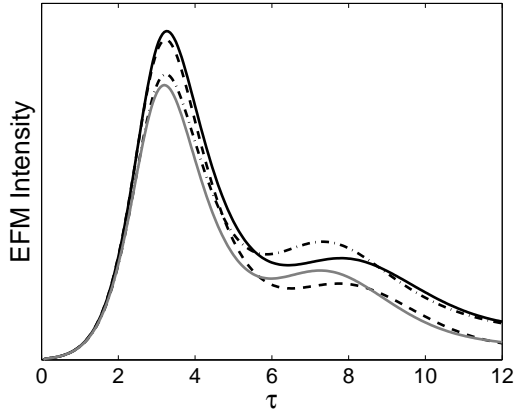}
\caption{Comparison of EFM intensity with $\gamma/\delta_L = 8\times 10^{-3}$, $\aleph=1.7$ for superradiance only
(black solid
line), superradiance and photon recoil (dashed line), superradiance and mean-field interaction (dot-dashed line), and
superradiance with both mean-field interactions and photon recoil included (gray line).
}\label{efmaav}
\end{center}
\end{figure}
\begin{figure}[t]
\centering
\begin{center}
\includegraphics[angle=0, scale=0.5]{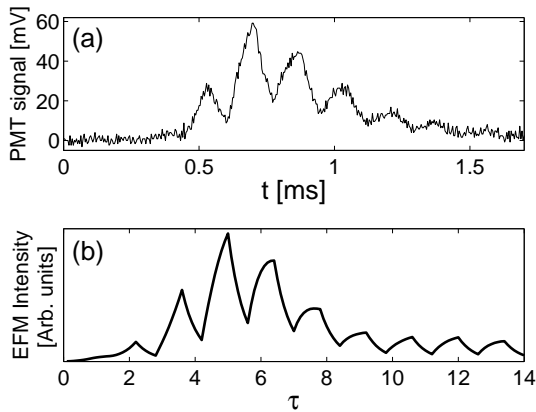}
\caption{Comparison of (a) experimental photomultiplier signal of EFM intensity to, see \cite{Stamperkurn2006}, (b)
a theoretical simulation with $\aleph=1.7$.  Experimental data by D. Stamper-Kurn \textit{et
al.} \cite{privcom2006}.}\label{efmdat}
\end{center}
\end{figure}

The semi-classical model is valid provided the relevant modes are macroscopically occupied. In the above simulations the
initial seed amplitude of the recoiling atomic modes was treated as a free parameter that may in general be chosen to
fit experimental data.  To study the effects of this choice of initial amplitude we plot in Fig.~\ref{ampfig} the time
evolution of EFM intensity for initial densities of the recoiling modes being respectively  $10^{-6}$ (solid black
line),  $10^{-5}$ (dashed line) and  $10^{-4}$ (gray line) times smaller than the condensate density.  For smaller
initial amplitude the time to peak superradiance is delayed and the ringing peak is higher relative to the intensity
maximum.
\begin{figure*}[t]
    \includegraphics[height=0.3\textheight, width=0.28\textwidth, draft=false, clip=true]{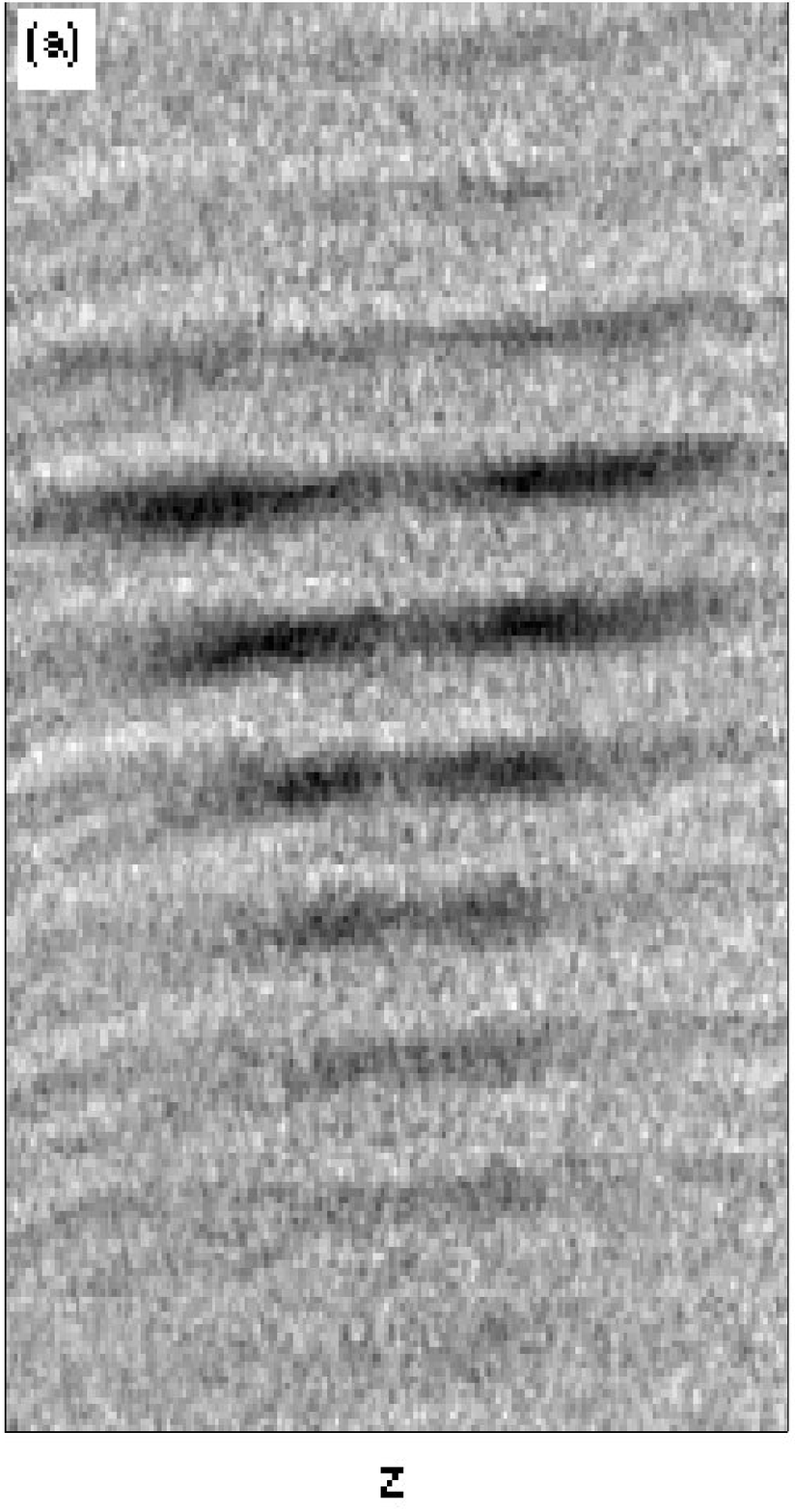}
	\includegraphics[height=0.3\textheight, width=0.32\textwidth,, draft=false, clip=true]{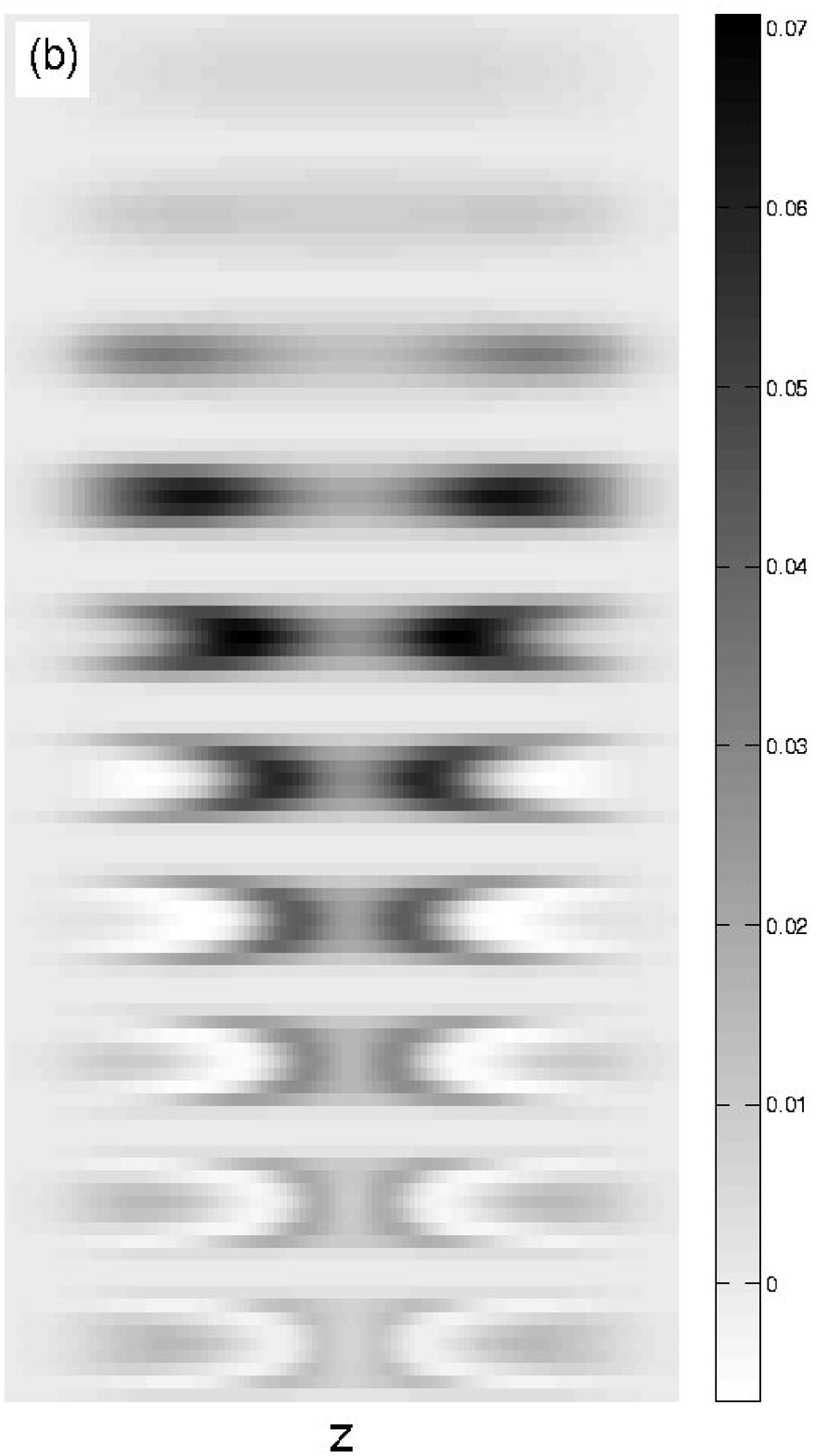}
    \includegraphics[height=0.3\textheight, width=0.28\textwidth, draft=false, clip=true]{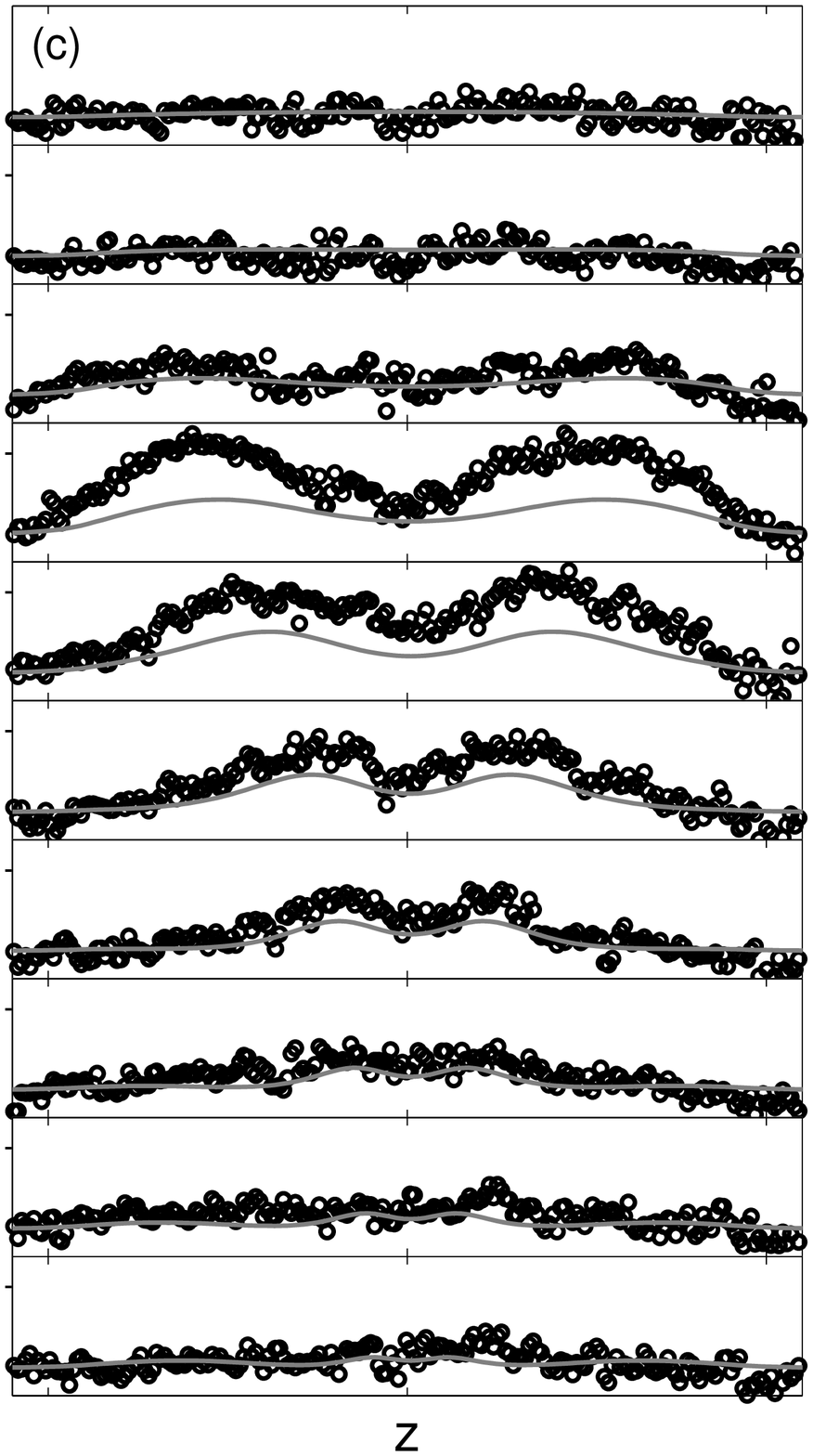}
	\caption{(a) Experimental coherent absorption imaging data, see \cite{Stamperkurn2006}, in which darkest regions
indicate strongest absorption, with a maximum absorption of $\sim 15\%$. (b) Simulated absorption images for
$\aleph=1.7$ and an initial density of
recoiling modes $|\Psi_0|^2/10^4$. (c) Comparison of experimental (circles) to simulated (gray lines) transverse
averaged absorption profiles. The maximum in the experimental data is $\sim 6\%$. In all figures time
evolves from top to bottom and the horizontal frame width is $L\approx 200\mu m$.  Experimental data by D. Stamper-Kurn
\textit{et
al.} \cite{privcom2006}.}\label{absdat}
\end{figure*}
    \begin{figure}[t]
    \centering
    \begin{center}
    \includegraphics[angle=0, scale=0.4]{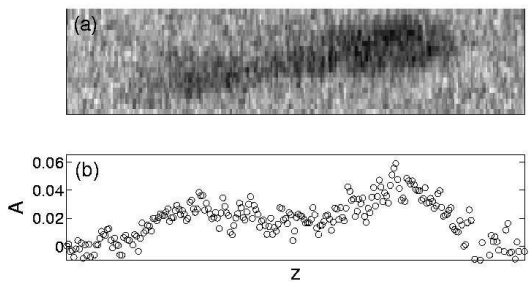}
	\caption{Example of (a) Typical absorption image, see \cite{Stamperkurn2006},  and (b) its averaged profile,
demonstrating asymmetry that
may arise from quantum fluctuations.  Experimental data by D. Stamper-Kurn \textit{et
al.} \cite{privcom2006}.}
	\label{asmdata}
    \end{center}
	\end{figure}

Now consider the effects of mean-field interactions and photon recoil.  Figure \ref{efmaav} shows the time
evolution
of the EFM intensity for $\gamma/\delta_L = 8\times 10^{-3}$ and $\aleph=1.7$. The results are for simulations
taking into
account superradiance only (solid
black line), superradiance and recoil velocity (dashed line), superradiance and mean-field interaction (dot-dashed
line), and superradiance with both mean-field interactions and recoil velocity (gray line).  The recoil velocity was so
chosen that an atom would
traverse roughly half a condensate width over $\Delta\tau=12.0$.  The primary effect of including the photon recoil
and mean-field interactions is to reduce the EFM amplitude.  Note that the height of the ringing peak is increased
somewhat relative to the first intensity maximum when including only mean-field interactions.  But generally speaking we
do not find significant qualitative changes in the spatial and temporal evolution of either
matter or light fields other than a reduction in overall amplitudes and slight relative delays.

\subsection{Comparison with experiment}
 
The primary purpose of the multimode treatment is to describe the quantum fluctuations that govern the initial
evolution of the system. As discussed in Sec. \ref{QF} this entails randomly choosing seed amplitudes for the intial
state of the recoiling atomic modes with the appropriate statistical properties. We have found, however, that the
multimode description gives the same results as a single-mode model as far as EFM intensity and absorption profiles are
concerned, provided that the sum of initial seed densities in the multimode case is equal to the seed density of the
single-mode system. For a single run of the experiment a single-mode description is therefore sufficient if the seed
densities are chosen judiciously. We thus turn to a comparison of the single-mode theory to data obtained in experiments
at UC Berkeley \cite{privcom2006}.
\begin{figure*}[t]
    \centering
    \begin{center}
    \includegraphics[angle=0, scale=0.55]{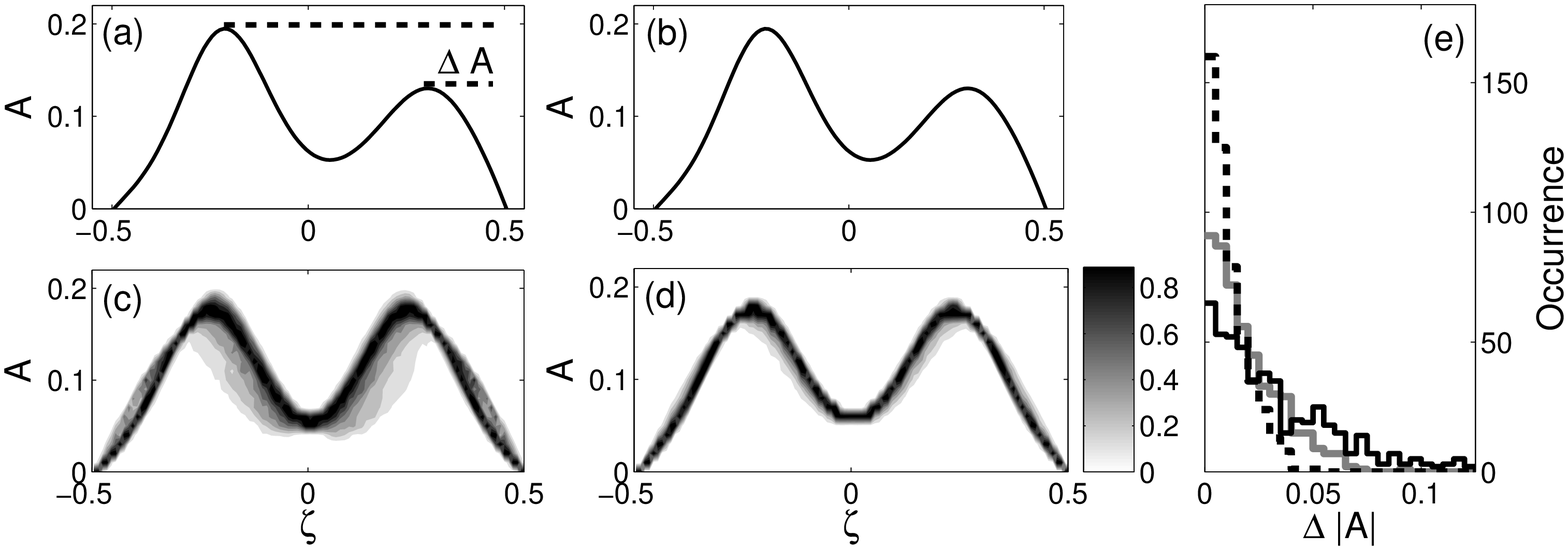}
    \end{center}
\caption{Absorption profile $A(z)$ of the input laser field.
Figures (a)-(b) are representative profiles resulting from single
realizations of a superradiance experiment for geometries with (a)
$m=5$ and (b) $m=20$ quasi-modes each in the left- and right EFM
directions. The peak separation, $\Delta\alpha$, as indicated by
the dotted lines in (a) characterizes the asymmetry. Figures (c)
and (d) give the frequency of occurrence of the absorption
profiles resulting from 500 realizations for  (c) $m=5$ and (d)
$m=20$. As indicated by the gray scale in the color bar, the
darkest shade of gray corresponds to the most frequent occurrences.
(e) Distribution of the asymmetry of absorption peaks from
500 realizations, for systems with $m=5$ (black line), $m=10$
(gray line) and $m=20$ (dashed line) quasi-modes in both the
left- and right EFM directions.}\label{absdist}
\end{figure*}

As described in Ref. \cite{Stamperkurn2006}, a Bose condensate of $N=1.6\times10^6$ $^{87}$Rb atoms trapped in an
Ioffe-Pritchard trap
with trap frequencies of $\omega_{x,y,z}=2\pi(48,48,5)s^{-1}$ was superradiantly pumped via the $F=1\rightarrow
F^\prime=1$ D1 transition from the $|F=1,m_f=-1\rangle$ to the $|F=2,m_f=1\rangle$ groundstates via a sequence of ten
pump pulses of $100\; \mu s$, each separated by a $68\;\mu s$ delay.  Fig. \ref{efmdat}(a) shows a typical EFM pulse
sequence recorded on a photomultiplier tube (PMT).  The rise time between pulses is a consequence of the slow response
time of the PMT. The lack of ringing indicates that dissipation plays a significant role in the dynamics.  Dissipation
due to spontaneous emission may be estimated using
$\gamma/\delta_L=\omega^3d^2/3\pi\epsilon_0\hbar c^3\delta_L\approx 8\times 10^{-3}$ \cite{Metcalf2002}.  Given an
$s$-wave
scattering length for $^{87}$Rb of $a\approx 10^2a_0$ where $a_0$ is the Bohr radius, we find $\aleph\approx 1.7$. The
experimental data agree qualitatively with the three dimensional simulation, see Fig. \ref{efmdat}(b). Here the initial
density of recoiling atoms was chosen $10^4$ times smaller than the condensate density, and the
dimensionless time scale was such that an atom would traverse roughly half a condensate width over the course of
the pulse sequence as in the Berkeley experiments. The PMT response was accounted for by convoluting the EFM intensity
with a response function with response time of the order of the pulse separation. 

Experimental absorption images corresponding to the PMT signal of Fig. \ref{efmdat}(a) are shown in Fig.
\ref{absdat}(a), where time evolves from top to bottom and each image is the time averaged absorption over the duration
of each pulse in the sequence.
Consistently with the observation that in this experimental run the asymmetry in the
absorption profile is small, we have taken the initial amplitudes of the atomic side modes $\Psi_{\pm k}$ to be
equal. 
The resulting absorption images are shown in Fig. \ref{absdat}(b).   An effect
not resolved in Ref. \cite{Stamperkurn2006} is the weak absorption seen at the upper- and lower edges
of the condensate in the lower six images of Fig. \ref{absdat}(b).  This absorption arises since the edges of the
condensate are less dense than the center, thus delaying superradiance in time and intensity.
 A comparison of our simulation to transverse averaged experimental absorption profiles is shown in Fig.
\ref{absdat}(c). The qualitative agreement is quite satisfactory although the value of the peak absorption is
underestimated by a factor of $\sim 2$.  Note that the averaging procedure supresses gain and other secondary features
seen in the simulation of Fig. \ref{absdat}(b), thus leading to the simple double peaked structure of Fig.
\ref{absdat}(c).
The simulation furthermore suggests that these features are too small in spatial extent and amplitude to be resolved in
the experimental data.  We emphasize that the only free parameters in this simulation are the initial seed amplitude
and the total time duration of the pulse sequence.  They are tightly constrained by, respectively,  the rise time to
peak superradiance relative to the EFM pulse width and the EFM pulse width relative to the overall time duration of the
pulse sequence.

\subsection{Shot-to-shot fluctuations}

We now turn to the shot-to-shot fluctuations resulting from the
quantum noise-dominated early stages of the emission process.  In Fig. \ref{asmdata}(a) we show a typical absorption
image obtained experimentally and exhibiting  strong asymmetry, as well as its transverse averaged profile in Fig.
\ref{asmdata}(b).
To account for this asymmetry we solve the multimode Eqs.~(\ref{psi1})-(\ref{psi2}) and
(\ref{propagate1})-(\ref{propagate2}) for $\aleph=3.0$, neglecting mean-field interactions
and effects due to photon recoil, with the initial amplitude
of each quasi-mode chosen from the probability distribution
(\ref{pdist}) independently of the other modes. The average
particle number $\bar n_\mb{q}$ is taken to be the same for all
distributions.

Figures \ref{absdist}(a)-(b) show absorption profiles for two randomly
chosen realizations of a superradiance experiment for systems with
(a) $m=5$ and (b) $m=20$ quasi-modes each in the left- and right
EFM directions of propagation.  In all cases the absorption profile is shown at the time of the
first EFM intensity maximum in either the left- or right traveling EFM. The separation $\Delta A$, as
indicated by the dotted lines in (a), may be used to characterize
the asymmetry.
Figure \ref{absdist}(c)-(d) shows a gray scale plot constructed
from 500 realizations of the superradiance experiment for (c)
$m=5$ and (d) $m=20$ quasi-modes each in the left- and right
travelling EFMs. Each profile is again taken at the first EFM intensity maximum.  The gray scale represents the
frequency of occurrence of absorption, with darkest shade of gray being the
highest frequency. 

The distribution of absorption asymmetries, $\Delta A$, for a
random set of 500 realizations of the experiment is plotted in
Fig. \ref{absdist}(e) for $m=5$ (black line), $m=10$ (gray line) and
$m=20$ (dashed line) quasi-modes each in the left- and right EFM
directions.  As may be expected the width of the distribution decreases with
increasing number of modes.

\section{Conclusion \label{summ}}

We have presented a multimode model of Raman superradiance scattering
to explain recent experiments that image Bose-condensates based on the abovementioned principle.  The superradiant
scattering into end-fire-modes is seen to lead to non-trivial, time-dependent,
spatial structure in both matter- and photon fields. The
microscopic quantum fluctuations that trigger the initial
superradiant dynamics lead to macroscopic fluctuations and
asymmetries in the spatial features of absorption images during
later stages.  The modeled absorption profiles are in
good qualitative agreement with experimental observations, confirming that superradiant scattering can be used to probe
the condensed phase of a BEC at finite temperature while remaining
insensitive to the non-condensed phase.

\section*{Acknowledgments}
We thank D. Stamper-Kurn for making his experimental results
available prior to publication, and D. Meiser for countless usefull discussions on this
work.

This work is supported in part by the US Army Research Office,
NASA, the National Science Foundation and the US Office of Naval
Research.

\renewcommand{\theequation}{A-\arabic{equation}}
\setcounter{equation}{0}  
\section*{APPENDIX A}  

The short-time expressions may be obtained by making an approximation of zero depletion for the laser field,
$\mathcal{E}_L\approx1$, and keeping only highest order terms in the equations of motion for the atomic fields and
end-fire-modes,
\begin{eqnarray}
\frac{d \Psi_0(\mathbf{r},t)}{d\tau} &\approx& \frac{i}{(1-i\gamma/\delta)}|\mathcal{E}_L|^2\Psi_0
\label{Apsi1}\\
\frac{d\Psi_\mb{k}(\mathbf{r},t)}{d \tau} &\approx&
\frac{i}{(1-i\gamma/\delta)}\mathcal{E}_\mb{-k}^*\mathcal{E}_L{\Psi}_0
 \label{Apsi2}\\
{\rm sign}(k) \frac{\partial\mathcal{E}_k}{\partial \zeta}
&\approx&\frac{i\aleph}{(1-i\gamma/\delta)}\Psi_{-k}^\dagger\mathcal{E}_L\Psi_0\label{Apropagate2}
\end{eqnarray}
Equation (\ref{Apsi1}) may immediately be solved to give 
\begin{equation}
\label{Apsi1zeroth}
\Psi_0(\tau) = \Psi^{(0)}_0e^{i\eta\tau}e^{-\nu\tau}
\end{equation}
where $\eta = \frac{1}{(1+(\gamma/\delta)^2)}$,  $\nu = \frac{\gamma/\delta}{(1+(\gamma/\delta)^2)}$ and
$\Psi^{(0)}_0$ is the initial condensate wave function which is assumed to be of the form
$\Psi^{(0)}_0=\phi_\zeta\phi_{(\xi,\Upsilon)}$.  Here $\Upsilon = y/L$ is the
dimensionless length in the
$y$-direction. We also assume that in the initial state the recoiling
atomic modes have the same wavefunction as the condensate, but smaller in amplitude by a factor $\beta\ll1$.
At time $\tau=0$ the EFM field is, by direct integration of Eq.~(\ref{Apropagate2}),
\begin{eqnarray}
\label{EFM0a}
{\rm sign}(k)\mathcal{E}_k(\zeta)&=&\aleph(i\eta-\nu)\beta|\phi_{(\xi,\Upsilon)}|^2\left(\int_{-\frac{1}{2}}^{\zeta}
|\phi_{\zeta^\prime}|^2d\zeta^\prime+\mathcal{E}_0\right)\nonumber\\
&=& \aleph(i\eta-\nu)\beta|\phi_{(\xi,\Upsilon)}|^2\Delta(\zeta)/\alpha,\label{EFM0b}
\end{eqnarray}
where $\mathcal{E}_0$ is chosen so that the boundary condition $\mathcal{E}_{-k}(1/2)=0$ is satisfied and
\begin{equation}
\Delta(\zeta)=\alpha\left(\int_{-\frac{1}{2}}^{\zeta}|\phi_{\zeta}|^2d\zeta^\prime + \mathcal{E}_0\right)
\end{equation}
where
\begin{equation}
\alpha=\frac{\aleph}{(\left(1+(\gamma/\delta)^2\right)}\left|\phi_{\xi,\Upsilon}\right|^2.
\end{equation}
Now substitute Eq. (\ref{Apsi1zeroth}) into Eq. (\ref{Apropagate2}) and define
$\tilde\mathcal{E}_{-k}^* = \mathcal{E}_{-k}^*e^{i\eta\tau+\nu\tau}$.  Then upon taking the time derivative of Eq.
(\ref{Apropagate2}) and
substitution of Eq. (\ref{Apsi2}) therein we find
\begin{equation}
\label{DEtz}
{\rm sign}(k)\frac{\partial^2\tilde\mathcal{E}_{-k}^*}{\partial\tau\partial\zeta}
=\alpha
\left|\phi_\zeta\right|^2\tilde\mathcal{E}_{-k}^*e^{-2\nu\tau}.
\end{equation}
To solve this differential equation we first treat the case of zero dissipation, \textit{i.e.} $\eta\rightarrow 1$ and
$\nu\rightarrow 0$, using the method of Laplace Transforms. Taking the Laplace transform with respect to time of
Eq. (\ref{DEtz}) we obtain
\begin{equation}
\label{DELapl}
s\frac{\mathsf{E}_{-k}^*(\zeta,s)}{\partial\zeta} - i\aleph\beta\big|\Psi^{(0)}_0\big|^2 =
\alpha|\phi_{\zeta}|^2\mathsf{E}_{-k}^*(\zeta,s),
\end{equation}
in which $\mathsf{E}_{-k}^*(\zeta,s)=\mathcal{L}\left(\tilde\mathcal{E}_{-k}^*\right)$. Equation (\ref{DELapl}) is a
linear first order differential equation and my be solved by standard means. We find
\begin{equation}
\mathsf{E}_{-k}^*(\zeta,s)= be^{\Delta(\zeta)/s} - i\aleph\beta|\phi_{(\xi,\Upsilon)}|^2/\alpha.
\end{equation}
The EFM field is found by inversion of the Laplace transform
\begin{equation}
\label{invLapl}
\mathcal{E}_{-k}^*(\zeta,\tau) = \frac{1}{2\pi
i}\int_{\mu-i\infty}^{\mu+i\infty}\;e^{s\tau}\mathsf{E}_{-k}^*(\zeta,s)ds.
\end{equation}
The contour of integration is chosen so as to include all poles of the integrand and the integral may be evaluated using
the residue theorem.  This is done by expanding both exponentials in the integrand in series and multiplying
out the result.  Integrating term-by-term the residue is found to be of the form
\begin{equation}
\sum\limits_{n=1}^\infty \frac{\tau^{n-1}\Delta^n}{(n-1)!n!} =
\sqrt{\frac{\Delta}{t}}I_1\left(2\sqrt{\tau\Delta}\right),
\end{equation}
where $I_q$ is a modified Bessel function of the first kind of order $q$. Thus
\begin{equation}
\label{EFMnu0}
\mathcal{E}_{-k}^*(\zeta,\tau) = i\beta\sqrt{\frac{\Delta}{\tau}}I_1\left(2\sqrt{\tau\Delta}\right)\Big|_{\nu=0},
\end{equation}
and by substitution into Eq. (\ref{Apropagate2})
\begin{equation}
\Psi_\mb{k}(\mathbf{r},\tau) = \beta\Psi^{(0)}_0I_0\left(2\sqrt{\tau\Delta}\right)\Big|_{\nu=0}.
\end{equation}

We now turn to the case with non-zero dissipation. Since the solution to the latter must give rise to Eq.
(\ref{EFMnu0}) for $\nu\rightarrow0$, we suggest a solution of the form
\begin{equation}
\label{EFMexp}
\mathcal{E}_{-k}^* \sim a_1(\tau)\Delta + a_2(\tau)\Delta^2 + a_3(\tau)\Delta^3 + ...
\end{equation}
were the $a_n(\tau)$ are time dependent coefficients that must vanish at $\tau = 0$ and reduce to
$\frac{\tau^{n-1}}{(n-1)!n!}$ for $\nu\rightarrow0$.  Substituting Eq. (\ref{EFMexp}) into Eq. (\ref{DEtz}) one obtains
a recursion relation for the $n$th coefficient 
\begin{equation}
\frac{\partial a_n}{\partial \tau} = \frac{a_{n-1}}{n}e^{-2\nu\tau}.
\end{equation}
To satisfy the $\nu\rightarrow 0$ condition we find that
\begin{equation}
a_n = \frac{1}{2^{n-1}\nu^{n-1}(n-1)!n!}\left(1-e^{-2\nu\tau}\right).
\end{equation}
The latter leads to the desired solutions
\be
\label{efmsln}
\tilde\mathcal{E}_{-k}^*(\mathbf{r},\tau) = i\beta(1-i\gamma/\delta)
\left(\frac{2\nu\Delta(\zeta)}{1-e^{-2\nu\tau}}\right)^{\frac{1}{2}}I_1\left(g(\zeta)\right),
\ee
and
\begin{equation}
\label{psisln}
\Psi_{k}(\mathbf{r},\tau) = \beta\Psi^{(0)}_0I_0\left(g(\zeta)\right).
\end{equation}
The argument of the Bessel functions is given by
\be
g(\zeta)=\left(\frac{2\Delta(\zeta)(1-e^{-2\nu\tau})}{\nu}\right)^{\frac{1}{2}}.
\ee

Finally, to find absorption profiles of the laser field we have to include all terms in the differential equation
\begin{equation} \frac{\partial \mathcal{E}_L}{\partial
\xi}=\frac{i\aleph}{(1-i\gamma/\delta)}\Psi_0^\dagger\left\{\mathcal{E}_L\Psi_0
+\sum_k\mathcal{E}_k\Psi_{-k}\right\}.
\end{equation}
For simplicity we take a uniform distribution in the $\xi$-direction since the exact
$\xi$-profile should not be too important in the short time limit. Substituting the short-time solutions obtained above
we find
\bwt
\begin{equation}
\mathcal{E}_L(\zeta) = e^{f(\zeta)} -\sum\limits_{\sigma=\pm1}
\aleph\beta\big|\Psi_0^{(0)}\big|^2\frac{(1+i\gamma/\delta)}{(1-i\gamma/\delta)}\frac{2\Delta(\zeta)}{
g(\sigma\zeta) } I_1\left(g(\sigma\zeta)\right)I_0\left(g(\sigma\zeta)\right)\frac{e^{f(\zeta)}-1}{f(\zeta)},
\end{equation}
\ewt
where
\begin{equation}
f(\zeta) = \aleph\left(i\eta-\nu\right)\phi_\zeta\int_{-\frac{w}{2L}}^{\frac{w}{2L}}\phi_{(\xi,\Upsilon)}d\xi.
\end{equation}
For a Thomas-Fermi profile $\phi_\zeta = \sqrt{(1/2)^2-\zeta^2}$ and 
\be
\Delta(\zeta) = \alpha\left(\frac{2}{3}(1/2)^3-(1/2)^2\zeta+\frac{1}{3}\zeta^3\right).
\ee

\bibliography{super}

\end{document}